\documentclass[aps,prl,reprint,superscriptaddress,longbibliography]{revtex4-2}

\usepackage{amsmath,amssymb,bm}
\usepackage{graphicx}
\usepackage{hyperref}
\usepackage{siunitx}
\begin{document}

\title{Giant second-harmonic generation in few-atomic layer metals}

\author{Alessio Zaccone}
\email{alessio.zaccone@unimi.it}
\affiliation{Department of Physics ``A. Pontremoli'', University of Milan, via Celoria 16, 20133 Milan, Italy}

\date{\today}

\begin{abstract}
Quantum confinement restructures the electronic phase space of ultrathin
metals, but the resulting density-of-states (DOS) change is not generically
large. We show that, within a three-dimensional confinement theory based on
the suppression of long-wavelength electronic states, the Fermi-level DOS
enhancement on the weak-confinement branch is bounded by
$(4/3)^{1/3}-1\simeq10.1\%$. Instead, confinement-induced redistribution of electronic
phase space can be strongly amplified by resonant denominators in the
second-order susceptibility. A minimal double-resonance closure predicts giant amplification when the film thickness approaches
$L_c=(2\pi/n)^{1/3}$, where $n$ is the carrier density, and a dominant optical pathway lies within a dephasing
linewidth of resonance. Applied to recent Ag(111) measurements, the theory
gives an effective phase-space carrier density parameter
$n_{\rm eff}\simeq3.6\times10^{20}\,\mathrm{cm}^{-3}$, a tenfold enhancement
of $|\chi^{(2)}|$, and hence the observed $\sim100$-fold increase of
thickness-normalized second-harmonic-generation efficiency, together with the observed non-oscillatory thickness dependence.
\end{abstract}

\maketitle

Quantum confinement in metallic films is commonly described through discrete
quantum-well states (QWS), with free in-plane motion and quantization along the
film normal \cite{Chiang2000}. For a real few-layer metal, however, the system
remains three dimensional and only one spatial direction is geometrically
confined. Atomic-scale variations (atomic "roughness") of the two interfaces can make the local
confinement length a function $L(x,y)$, so that $k_z$ need not remain an exact
good quantum number even when QWS remain spectroscopically visible
\cite{Zaccone2025}. This distinction motivates a complementary phase-space
description in which confinement removes a finite region of the
three-dimensional momentum manifold rather than replacing the out-of-plane
degree of freedom solely by an ideal hard-wall ladder.

The confinement theory of
Refs.~\cite{Travaglino2023,Zaccone2025,Ummarino_2025} implements precisely this
construction and has already yielded quantitative, parameter-free agreement
with superconducting thin-film data within an Eliashberg treatment
\cite{Ummarino_2025}. Here we ask what the same phase-space restructuring
implies for nonlinear optics. The central result is twofold. First, the bare
Fermi-level DOS enhancement has a strict upper bound of only about $10\%$ on
the weak-confinement branch. Second, this modest redistribution can nevertheless
produce a giant nonlinear response when it is sampled by resonant denominators
of $\chi^{(2)}$. The theory therefore separates the origin of the spectral
redistribution from the mechanism that amplifies it.

\begin{figure*}[t]
    \centering
    \includegraphics[width=0.8\textwidth]{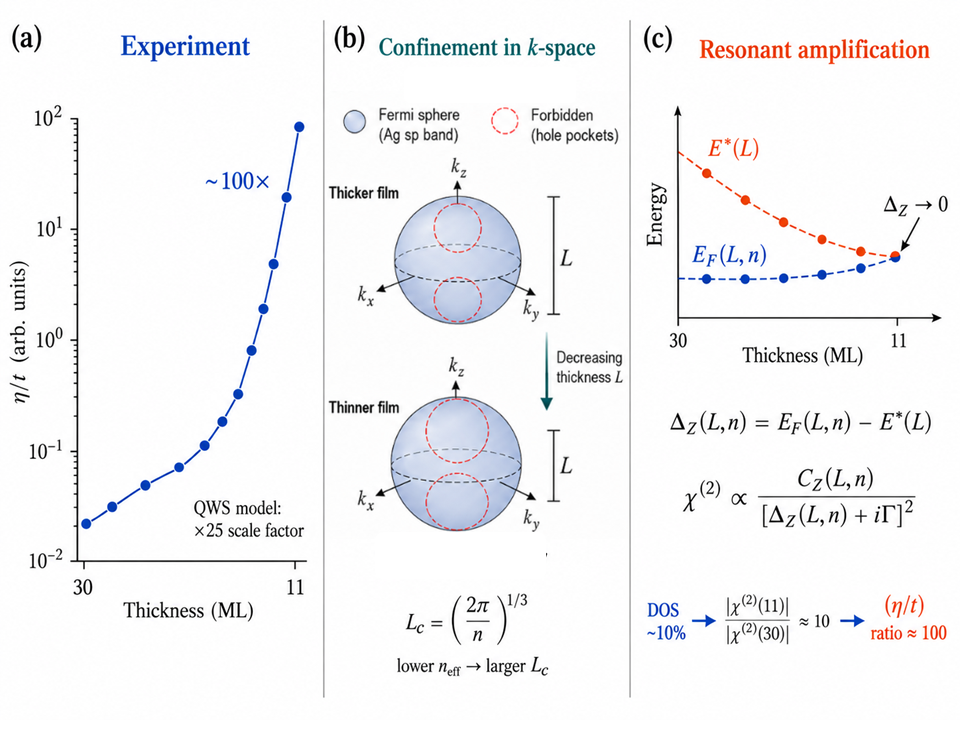}
    \caption{\textbf{Carrier-density-controlled confinement mechanism for
enhanced SHG in ultrathin Ag.}
(a) Schematic representation of the experimentally reported increase of the
thickness-normalized SHG efficiency by nearly two orders of magnitude between
30 and 11 ML. The QWS calculation of Ref.~\cite{Jenke2026} reproduces the
qualitative thickness trend, but its calculated $\chi^{(2)}$ was multiplied
by an overall factor of 25 for comparison with experiment.
(b) In the confinement picture \cite{Zaccone2025}, decreasing the film
thickness enlarges the two forbidden spherical regions of momentum space. A
reduced effective carrier-density parameter shifts the crossover
$L_c=(2\pi/n)^{1/3}$ toward the experimentally relevant thickness range.
(c) Minimal resonance closure used here. The confinement scale
$\Delta_Z(L,n)=E_F(L,n)-E^*(L)$ vanishes at $L_c$; if the dominant optical
detuning tracks this scale, resonant denominators amplify the otherwise modest
phase-space redistribution, yielding an order-of-magnitude increase in
$|\chi^{(2)}|$ and hence an approximately two-orders-of-magnitude increase in
$\eta/L$.}
    \label{fig:mechanism}
\end{figure*}

For a film confined along $z$ and extended in the $xy$ plane, the maximum
wavelength propagating at polar angle $\theta$ is \cite{Zaccone2025}
\begin{equation}
\lambda_{\max}=\frac{L}{\cos\theta},
\qquad
k_{\min}=\frac{2\pi\cos\theta}{L}.
\label{eq:kmin}
\end{equation}
The resulting allowed $k$-space volume in the weak-confinement regime is the
Fermi sphere minus two spherical pockets of radius $\pi/L$,
\begin{equation}
\mathcal V_k=\frac{4\pi}{3}k^3-\frac{8\pi}{3}\left(\frac{\pi}{L}\right)^3.
\label{eq:Vk}
\end{equation}
Including spin degeneracy, particle counting gives the free carrier density as
\begin{equation}
 n=\frac{2}{(2\pi)^3}\mathcal V_k
 =\frac{k_F^3}{3\pi^2}-\frac{2\pi}{3L^3},
\label{eq:ncount}
\end{equation}
so that the Fermi momentum follows as
\begin{equation}
 k_F=(3\pi^2n)^{1/3}
 \left(1+\frac{2\pi}{3nL^3}\right)^{1/3},
\label{eq:kF}
\end{equation}
and the Fermi energy as
\begin{align}
 E_F(L,n)&=E_F^{\infty}(n)
 \left(1+\frac{2\pi}{3nL^3}\right)^{2/3}, \nonumber\\
E_F^{\infty}&=\frac{\hbar^2}{2m^*}(3\pi^2 n)^{2/3}.
\label{eq:EF}
\end{align}
The confinement-modified electronic DOS crosses from the usual $\sqrt{E}$ law
to a low-energy linear law at \cite{Travaglino2023,Zaccone2025}
\begin{equation}
 E^*(L)=\frac{2\pi^2\hbar^2}{m^*L^2}.
\label{eq:Estar}
\end{equation}
The condition $E_F=E^*$ gives the crossover thickness
\begin{equation}
 L_c=\left(\frac{2\pi}{n}\right)^{1/3}.
\label{eq:Lc}
\end{equation}
For $L>L_c$, the Fermi surface remains topologically spherical although the
two forbidden pockets (cf. Fig. \ref{fig:mechanism}(b)) grow upon reducing $L$; at $L_c$ the
confinement-induced topological crossover of
Refs.~\cite{Travaglino2023,Zaccone2025} is reached.

A general no-go result follows directly from the DOS. In the $L>L_c$ branch,
evaluating the square-root DOS at Eq.~\eqref{eq:EF} gives
\begin{equation}
 C_Z(L,n)\equiv
 \frac{g[E_F(L,n)]}{g_{\rm bulk}[E_F^{\infty}(n)]}
 =\left(1+\frac{2\pi}{3nL^3}\right)^{1/3}.
\label{eq:CZ}
\end{equation}
Writing $x=L/L_c$,
\begin{equation}
 C_Z(x)=\left(1+\frac{1}{3x^3}\right)^{1/3},
\qquad x\ge 1,
\label{eq:Cx}
\end{equation}
and therefore
\begin{equation}
 C_Z(L_c)=\left(\frac{4}{3}\right)^{1/3}=1.1006.
\label{eq:Cmax}
\end{equation}
Thus, within this confinement branch, a DOS-only mechanism can enhance the
Fermi-level DOS by at most $10.1\%$ and cannot by itself generate an
order-of-magnitude nonlinear susceptibility. This contrasts with
superconductivity, where the same DOS modification enters exponentially
through $T_c\propto\exp[-1/Ug(E_F)]$
\cite{Travaglino2023,Zaccone2025}.

The required amplification is naturally available in nonlinear optics. In a
centrosymmetric metal the bulk electric-dipole second-order response vanishes,
so that surface, interface, and nonlocal contributions dominate
\cite{Sipe1987,Boyd}. A generic independent-particle contribution to the
surface-normal susceptibility has the sum-over-states structure
\cite{Boyd,Echarri2021}
\begin{widetext}
\begin{equation}
\chi^{(2)}_{zzz}(2\omega;\omega,\omega)
\propto
\sum_{abc}
\int_{\mathcal K_{\parallel}(L,n)}
\frac{d^2k_{\parallel}}{(2\pi)^2}
\frac{
(f_a-f_b)z_{ab}z_{bc}z_{ca}
}{
[\hbar\omega-E_{ba}+i\Gamma_{ba}]
[2\hbar\omega-E_{ca}+i\Gamma_{ca}]
}
+\cdots .
\label{eq:chiSOS}
\end{equation}
\end{widetext}
Here $a,b,c$ label confinement-derived electronic states, or QWS subbands when
a QWS basis is used, at fixed in-plane wave vector $\mathbf{k}_{\parallel}$;
$f_a$ denotes the Fermi occupation,
$E_{ba}=E_b-E_a$ and $E_{ca}=E_c-E_a$ are transition energies,
$z_{ab}=\langle a|z|b\rangle$ are out-of-plane position matrix elements, and
$\Gamma_{ba}$ and $\Gamma_{ca}$ are phenomenological coherence linewidths.
The domain $\mathcal K_{\parallel}(L,n)$ denotes the in-plane projection of
the confinement-modified electronic phase space. The ellipsis represents the
remaining frequency permutations and intermediate-state contributions.
Equation~\eqref{eq:chiSOS} shows why a modest redistribution of phase space can
have a large effect when spectral weight is transferred through a one- or
two-photon resonance.

To expose this mechanism analytically, we introduce the confinement energy
scale
\begin{equation}
\Delta_Z(L,n)=E_F(L,n)-E^*(L),
\label{eq:Delta}
\end{equation}
which vanishes at $L=L_c$. We emphasize that $E^*(L)$ is a confinement
crossover scale and is not, by itself, a microscopic optical transition
energy. The minimal closure assumes only that, over a finite thickness
interval around the crossover, the dominant thickness-dependent part of a
relevant optical detuning follows the same scale $\Delta_Z(L,n)$. The
simplest double-resonance form containing no additional thickness-dependent
parameter is then
\begin{equation}
\chi_Z^{(2)}(L)=\chi_c^{(2)} C_Z(L,n)
\frac{\Gamma^2}{[\Delta_Z(L,n)+i\Gamma]^2}.
\label{eq:minimalchi}
\end{equation}
Taking the modulus of Eq.~(13) makes the resonant amplification explicit,
\begin{equation}
|\chi_Z^{(2)}(L)|
=
|\chi_c^{(2)}|\,C_Z(L,n)
\frac{\Gamma^2}{\Delta_Z^2(L,n)+\Gamma^2}.
\end{equation}
Thus the nonlinear response crosses from the off-resonant form
$|\chi^{(2)}|\propto\Gamma^2/\Delta_Z^2$ for
$|\Delta_Z|\gg\Gamma$ to its full resonant amplitude when
$|\Delta_Z|\lesssim\Gamma$. The giant enhancement is therefore generated
not by a divergent susceptibility, but by the ratio between an off-resonant
thick film and a nearly resonant film close to $L_c$. 
This phenomenological reduction of Eq.~\eqref{eq:chiSOS}  provides a direct test of
whether the confinement scale has the magnitude and thickness dependence
needed for resonant nonlinear amplification. The corresponding ratio is
\begin{equation}
\frac{|\chi_Z^{(2)}(L_1)|}{|\chi_Z^{(2)}(L_0)|}
=
\frac{C_Z(L_1,n)}{C_Z(L_0,n)}
\frac{\Delta_Z^2(L_0,n)+\Gamma^2}
{\Delta_Z^2(L_1,n)+\Gamma^2}.
\label{eq:Rchi}
\end{equation}
Equations~\eqref{eq:Lc}--\eqref{eq:Rchi} define a general screening criterion
for large confinement-enhanced nonlinear response: the film should lie near
$L_c$, where the phase-space redistribution is strongest on this branch, and
a microscopic optical pathway should simultaneously satisfy
$|\Delta_{\rm opt}|\lesssim\Gamma$. The first condition is set by confinement;
the second is spectroscopic. The two need not coincide in every material.

We now apply this framework to the recent measurements of Jenke
\textit{et al.} on crystalline Ag(111) films containing 10--30 atomic
monolayers (ML) \cite{Jenke2026}. At a pump wavelength
$\lambda_0=1.8\,\mu$m, the thickness-normalized SHG efficiency $\eta/L$
rises by almost two orders of magnitude between 30 and 11 ML. Remarkably, the
increase occurs without pronounced layer-by-layer oscillations, indicating
that a smooth contribution dominates over oscillatory QWS resonances or
thickness-dependent interference. The same experiment resolves QWS by ARPES
and compares the data with a microscopic QWS model derived from
Ref.~\cite{Echarri2021}. That calculation captures the increasing thickness
trend, but quantitative comparison requires multiplying the theoretical
$\chi^{(2)}$ by an overall factor of 25, while the calculated wavelength
dependence also differs from experiment \cite{Jenke2026}.

The physical connection between the general confinement mechanism and this
experiment is summarized in Fig.~\ref{fig:mechanism}.

The confinement construction is complementary to, rather than incompatible
with, the observed QWS. In the model used by Ref.~\cite{Jenke2026}, the
confining potential is translationally invariant parallel to the surface and
the electronic response is solved along the confinement direction, with the
in-plane dispersion incorporated through band-dependent effective masses.
This corresponds to an approximately separable quasi-two-dimensional
structure,
$\Psi_{\nu{\bf k}_{\parallel}}({\bf r})\sim
 e^{i{\bf k}_{\parallel}\cdot{\bf r}_{\parallel}}\phi_\nu(z)$.
The phase-space theory instead allows the electronic states contributing to
the nonlinear response to sample a genuinely three-dimensional momentum
manifold even while QWS remain spectroscopically well defined.

The Ag(111) layer spacing used in Ref.~\cite{Jenke2026} is
$a_{111}=2.36$~\AA. Thus $L_{11}=11a_{111}=2.596$ nm and
$L_{30}=7.08$ nm. If the confinement crossover lies near the lower end of the
measured interval, Eq.~\eqref{eq:Lc} immediately gives
\begin{equation}
 n_c\equiv\frac{2\pi}{L_{11}^3}
 =3.59\times10^{26}\,\mathrm{m}^{-3}
 =3.59\times10^{20}\,\mathrm{cm}^{-3}.
\label{eq:nc}
\end{equation}
This value is about $1/160$ of the conventional bulk free-electron density of
Ag. Because Eq.~\eqref{eq:Lc} follows from single-band phase-space counting,
$n_{\rm eff}$ should at this stage be regarded as an \emph{effective
phase-space carrier-density parameter} entering the confinement manifold,
rather than as a direct measurement of either the total Ag electron density
or the population of a specific QWS subband. Its possible microscopic
identification with a reduced delocalized QWS/$sp$ sector is a hypothesis to
be tested experimentally below.

For the Ag application we take $L_1=11$ ML and $L_0=30$ ML and fix
$\Gamma=55$ meV, representative of the tens-of-meV spectroscopic scale of
sharp Ag QWS/surface states \cite{Chiang2000,Nicolay2000}. Ref.~\cite{Jenke2026}
instead uses a homogeneous $0.1$ eV damping in its baseline calculation and
explicitly notes that microscopic linewidths are state and energy dependent;
the sensitivity to $\Gamma$ is reported in the Supplemental Material. The
experimental target
$|\chi^{(2)}(11\,\mathrm{ML})|/|\chi^{(2)}(30\,\mathrm{ML})|\simeq10$
follows from $\eta/L\propto|\chi^{(2)}|^2$ in the extraction used by
Ref.~\cite{Jenke2026}. Selecting the solution continuously connected to the
independently motivated condition $L_c\simeq11$ ML from Eq.~\eqref{eq:nc},
Eq.~\eqref{eq:Rchi} gives
\begin{equation}
 n_{\rm eff}=3.61\times10^{26}\,\mathrm{m}^{-3}
 =3.61\times10^{20}\,\mathrm{cm}^{-3},
\label{eq:nfit}
\end{equation}
corresponding to $L_c=10.98$ ML. For $m^*=m_e$, appropriate as a first
nearly-free-electron description of the Ag $sp$ band near $E_F$
\cite{AgSpBand}, one obtains
\begin{align}
\Delta_Z(11\,\mathrm{ML})&=0.00062~\mathrm{eV},\\
\Delta_Z(30\,\mathrm{ML})&=0.1569~\mathrm{eV},\\
\frac{C_Z(11)}{C_Z(30)}&=1.0942.
\end{align}
Insertion in Eq.~\eqref{eq:Rchi} gives
\begin{equation}
\frac{|\chi_Z^{(2)}(11)|}{|\chi_Z^{(2)}(30)|}=10.0,
\qquad
\frac{(\eta/L)_{11}}{(\eta/L)_{30}}\simeq100,
\label{eq:fitresult}
\end{equation}
in agreement with the reported magnitude at $1.8\,\mu$m
\cite{Jenke2026}. The same construction naturally produces a smooth thickness
envelope because the confinement-modified phase-space volume varies
continuously with $L$, rather than through a regularly spaced $k_z$ ladder.
Residual QWS oscillations can therefore modulate this envelope without
controlling its dominant thickness dependence, consistent with the absence of
pronounced oscillations in the experiment.

Is the inferred density scale plausible? Existing literature supports the
\emph{possibility} of strong reductions of a mobile or optically active
carrier density in nanoscale metals, but also shows that the effect is not
universal. Mirigliano \textit{et al.} found nonmetallic transport and carrier
localization in continuous nanostructured Au films, attributing the low
free-carrier population to grain boundaries, defects, Coulomb blockade, and
space-charge-limited conduction \cite{Mirigliano2020}. More directly, Das
\textit{et al.} measured by Hall effect a carrier concentration
$3.70\times10^{20}\,\mathrm{cm}^{-3}$ in a 2-nm epitaxial HfN film, nearly
identical to Eq.~\eqref{eq:nfit}, in a regime of confinement-induced plasmonic
breakdown \cite{Das2024}. For Ag itself, Gong \textit{et al.} found
thickness-dependent bulk-plasmon energies and interpreted the trend as an
increase of effective free-electron density toward the bulk value with
increasing thickness; their films below about 8 nm were optically nonmetallic
\cite{Gong2015}. Mendoza-Herrera \textit{et al.} likewise extracted a redshift
of the Ag and Au plasma frequency with decreasing thickness
\cite{Mendoza2022}.

The approximately 1-nm Si passivation layer of Ref.~\cite{Jenke2026}
oxidizes to SiO$_2$ upon ambient exposure; interfacial chemical modification
or defect-induced carrier trapping may therefore provide an additional
sample-specific route to a reduced $n_{\rm eff}$.

At the same time, high-quality ultrathin metals can retain bulk-like carrier
density. Akolzina \textit{et al.} found an essentially bulk-like plasma
frequency in continuous 3--50 nm Au films and confirmed a large carrier
density by Hall measurements \cite{Akolzina2024}. Moreover, ARPES on metallic
Ag(111)/H--Si(111) films found Fermi vectors close to single-crystal Ag
\cite{ArranzFS2002}, while well-defined Ag QWS are known in the same thickness
range \cite{ArranzQW2002}. 
The low-$n_{\rm eff}$ hypothesis is therefore
deliberately sample-specific.

Within the present reduced theory,
$n_{\rm eff}$ is most conservatively interpreted as the effective density
parameter controlling the accessible confinement phase space. Whether this
parameter corresponds microscopically to a reduced mobile density, a
restricted optically active QWS/$sp$ sector, or a combination of spectral
weight and carrier-density renormalization cannot be decided from the SHG
data alone.

The proposal is readily falsifiable. ARPES can count the occupied Fermi area
of the QWS; the fitted density corresponds at 11 ML to a sheet density
$n_{2D}=n_{\rm eff}L\simeq9.4\times10^{13}\,\mathrm{cm}^{-2}$, which for one
spin-degenerate circular pocket gives $k_F\simeq0.24$~\AA$^{-1}$, while $M$
equivalent pockets reduce this value by $M^{-1/2}$. Hall measurements directly
constrain the mobile density, and infrared ellipsometry or transmission can
extract the thickness-dependent Drude weight or plasma frequency. The
strongest theoretical test is to incorporate the confinement-modified phase
space directly into the existing microscopic QWS/RPA calculation
\cite{Echarri2021,Jenke2026}, replacing the unrestricted in-plane integration
by $\mathcal K_{\parallel}(L,n)$ and fitting all three excitation wavelengths
with the \emph{same} $n_{\rm eff}$. Failure of a common density parameter to
describe the three wavelengths would rule out the minimal closure.
Comparing different passivation schemes, or repeating the measurement in a
less oxidation-sensitive metal, would further distinguish intrinsic
confinement from sample-specific carrier trapping.

In conclusion, the three-dimensional confinement construction leads to a
general result for ultrathin-metal nonlinear optics: on the weak-confinement
branch the bare DOS enhancement is bounded to about $10\%$, so a giant
response requires an amplification mechanism. Resonant denominators in
$\chi^{(2)}$ provide such a mechanism when confinement brings the electronic
phase space near $L_c$ and an optically allowed pathway simultaneously lies
within a linewidth of resonance. The Ag(111) experiment provides a concrete
quantitative test: the minimal closure reproduces the observed magnitude and
predominantly non-oscillatory thickness dependence for
$n_{\rm eff}\sim3.6\times10^{20}\,\mathrm{cm}^{-3}$. Direct carrier counting,
optical Drude-weight measurements, and a microscopic QWS calculation using the
confinement-modified momentum manifold can now determine whether this
phase-space mechanism is realized in the measured films.

\begin{acknowledgments}
The author thanks the authors of Ref.~\cite{Jenke2026} for stimulating experimental results. Useful feedback from Prof. Paolo Milani is gratefully acknowledged. No new experimental data are reported here.
\end{acknowledgments}

\bibliography{zaccone_shg_prl}

\end{document}